\documentclass[12pt]{article}
\usepackage{epsfig,graphicx}
\usepackage{ch2005}

\def\beq{\begin{equation}}
\def\eeq#1{\label{#1}\end{equation}}
\def\eeqn{\end{equation}}

\def\beqa{\begin{eqnarray}}
\def\eeqa#1{\label{#1}\end{eqnarray}}
\def\eeqan{\end{eqnarray}}

\def\Dslash{\not{\hbox{\kern-4pt $D$}}}
\def\dslash{\not{\hbox{\kern-2pt $\del$}}}

\newcommand{\tev}{\ensuremath{\mathrm{\,Te\kern -0.1em V}}\xspace}
\newcommand{\gev}{\ensuremath{\mathrm{\,Ge\kern -0.1em V}}\xspace}
\newcommand{\mev}{\ensuremath{\mathrm{\,Me\kern -0.1em V}}\xspace}
\newcommand{\kev}{\ensuremath{\mathrm{\,ke\kern -0.1em V}}\xspace}
\newcommand{\ev}{\ensuremath{\mathrm{\,e\kern -0.1em V}}\xspace}
\newcommand{\gevc}{\ensuremath{{\mathrm{\,Ge\kern -0.1em V\!/}c}}\xspace}
\newcommand{\mevc}{\ensuremath{{\mathrm{\,Me\kern -0.1em V\!/}c}}\xspace}
\newcommand{\gevcc}{\ensuremath{{\mathrm{\,Ge\kern -0.1em V\!/}c^2}}\xspace}
\newcommand{\mevcc}{\ensuremath{{\mathrm{\,Me\kern -0.1em V\!/}c^2}}\xspace}
\def\mus  {\ensuremath{\rm \,\mus}\xspace}

\def\mus        {\ensuremath{\,\mu{\rm s}}\xspace}    %% microsecond

\begin{document}

%+ \Chapter{}
%+ {Shell-type Supernova Remnants}
%+ {Heinrich J.~V\"olk}

\Title{Shell-type Supernova Remnants}
\bigskip

%+ \addcontentsline{toc}{chapter}{{\it Heinrich J.~V\"olk}}
%+ \index{author}{V\"olk, H.J.} 

%%%%%%%%%%%%%%%%%%%%%%%%%%%%%%%%%%%%%
% Label to flag the first page of your contribution
% Replace Yourname by your name starting with a capital letter
%
\label{VoelkStart}

%%%%%%%%%%%%%%%%%%%%%%%%%%%%%%%%%%%%%
% Your name
%
\author{Heinrich J. V\"olk\index{V\"olk, H.J.}}

%%%%%%%%%%%%%%%%%%%%%%%%%%%%%%%%%%%%%
% Your address
%
\address{Max-Planck-Institut f\"ur Kernphysik\\
P.O. Box 103980 \\
D-69029 Heidelberg, Germany \\
}

\makeauthor\abstracts{ 
The role of Supernova Remnants (SNRs) for the production
  of the Galactic Cosmic Rays is reviewed from the point of view of theory and
  very high energy gamma-ray experiments. The point is made that theory can
  describe young SNRs very well, if the evidence from the synchrotron emission
  is used to empirically determine several parameters of the theory, and thus
  theory can predict the relative contributions of hadronic and leptonic
  gamma rays at TeV energies. This is exemplified for several objects that have
  been observed intensively during the last years. Future key observations are
  discussed.
}

\section{Introduction}
Shell-type supernova remnants
(SNRs) are widely assumed to be the sources of the Cosmic Rays (CRs), as they
are observed in the neighborhood of the Solar System. This concerns particle
energies up to the "knee" of the energy spectrum at several $10^{15}$~eV or
possibly beyond (see \cite{hillas05} for a recent review). From
estimates of the Galactic Supernova (SN) rate and the CR escape
rate from the Galaxy SNRs have then to convert on average about 10\% of their
entire mechanical explosion energy into CRs -- an enormous requirement.

A direct experimental investigation of SNRs as CR sources is possible with
$\gamma$-ray observations at very high energies $> 100$ GeV (VHE). The argument
is the following: the acceleration of particles to CR energies is assumed to
occur primarily at the outer, quasi-spherical shock which also compresses and
heats the ambient circumstellar medium. The expanding shock wave confines the
accelerated particles in its interior until its velocity decreases
substantially at late times. Then the shock gets ``old'' and the more energetic
\begin{figure}
\begin{minipage}[t]{0.47\linewidth}
\begin{center}
\epsfig{file=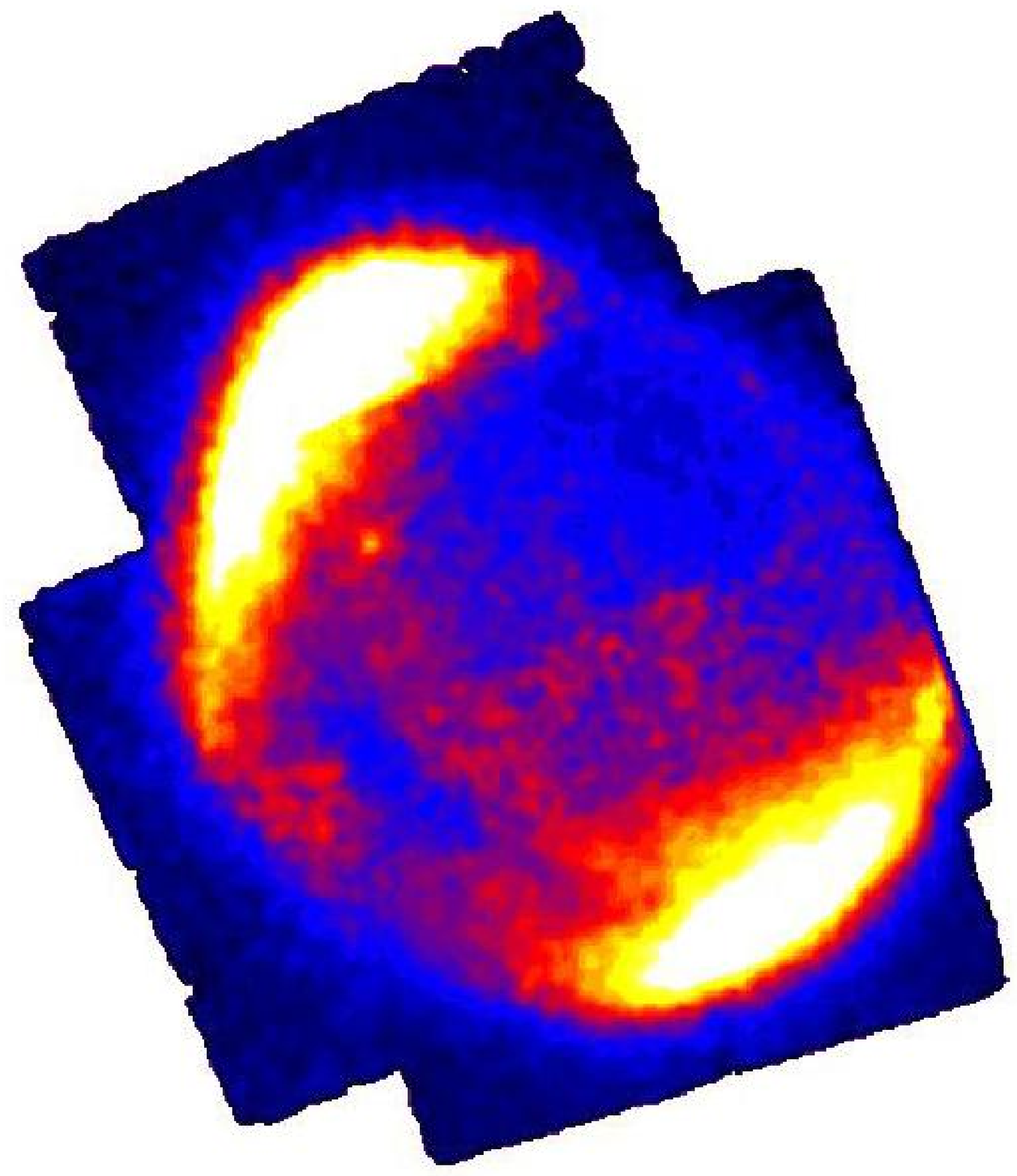,width=0.9\linewidth}
\caption{{\it ASCA} image of SN 1006 in hard X-rays from
\cite{koyama95}. The
emission comes mainly from the Northeastern and Southwestern areas, interpreted
as polar caps, where the mean magnetic field is quasi-parallel to the shock
normal of the outer SNR blast wave \cite{voelk97}. The resolution
of {\it ASCA} is slightly better than that of the {\it H.E.S.S.} array,
indicating what is presently achievable in VHE $\gamma$-rays.}
\label{fig:ASCA1006}
\end{center}
\end{minipage} 
\hspace{0.06\linewidth}
\begin{minipage}[t]{0.47\linewidth}
\begin{center}
\epsfig{file=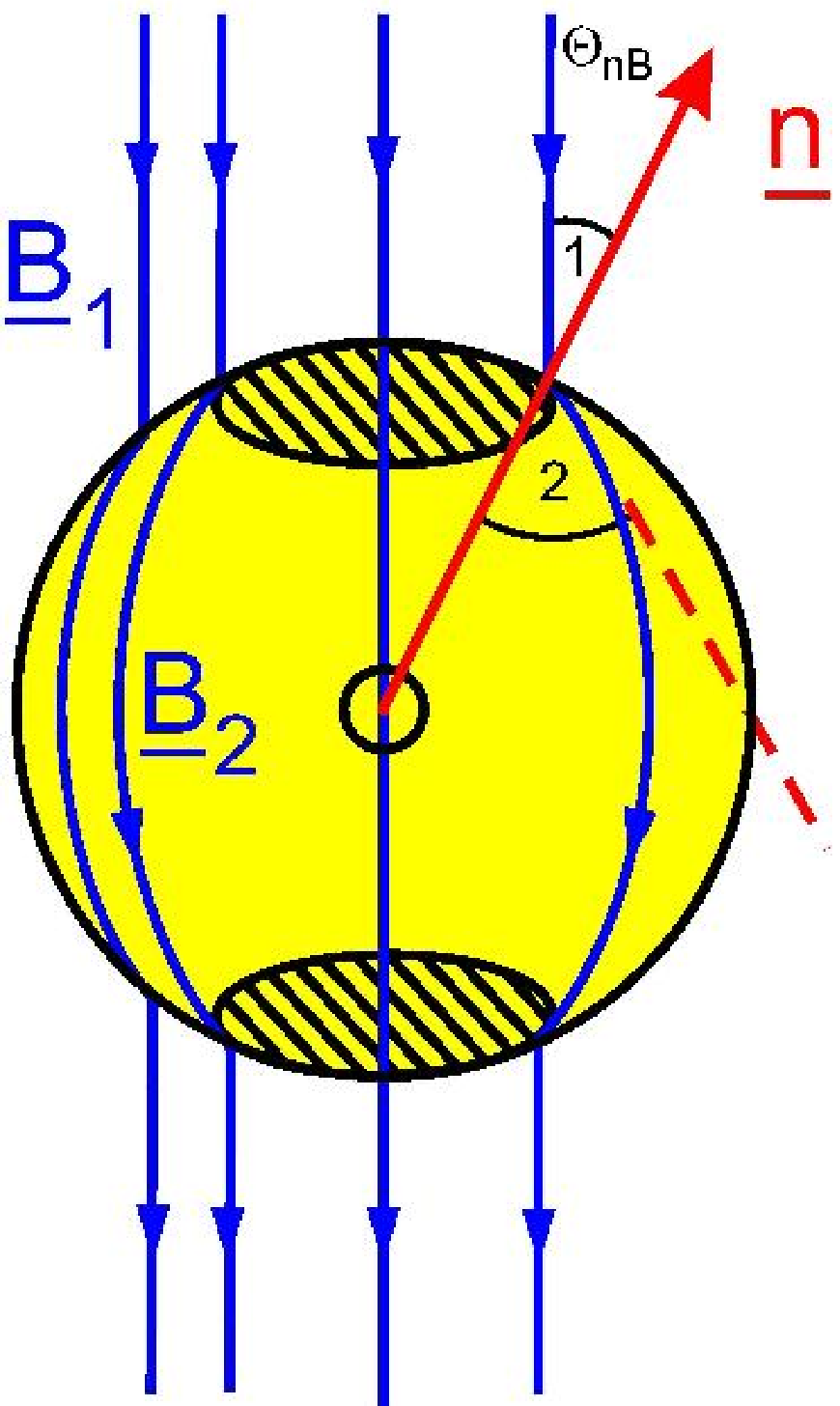,width=0.6\linewidth}
\caption{Schematic of the average magnetic field geometry for a SN explosion 
into a uniform interstellar medium with a homogenenous field
$\underline{B}_1$. $\Theta_{\mbox {nB}}$ is the angle between the shock normal
vector $\underline{n}$ and $\underline{B}_1$. Injection of downstream
suprathermal ion is only possible for sufficiently small values of
$\Theta_{\mbox {nB}}$, i.e. in the hatched polar regions.}
\label{fig:injection}
\end{center}
\end{minipage}
\end{figure}
particles successively leave the remnant. There fore a distant $\gamma$-ray
observer of a ``young'' SNRs can see the $\pi^0$-decay and nonthermal
Bremsstrahlung (NB) emission, due to CR collisions with gas particles in the
interior, jointly with the Inverse Compton (IC) radiation as originating from a
{\it localized} source. Together with the electron synchrotron spectrum -- from
radio to hard X-ray energies -- and the synchrotron morphology
(e.g. Fig.~\ref{fig:ASCA1006}) this is the nonthermal electromagnetic evidence. I
want to argue below that the synchrotron emission allows us to seperate the
contributions of nuclear CRs and ultrarelativistic CR electrons to the
\begin{figure}
\begin{minipage}[t]{0.47\linewidth}
\begin{center}
\epsfig{file=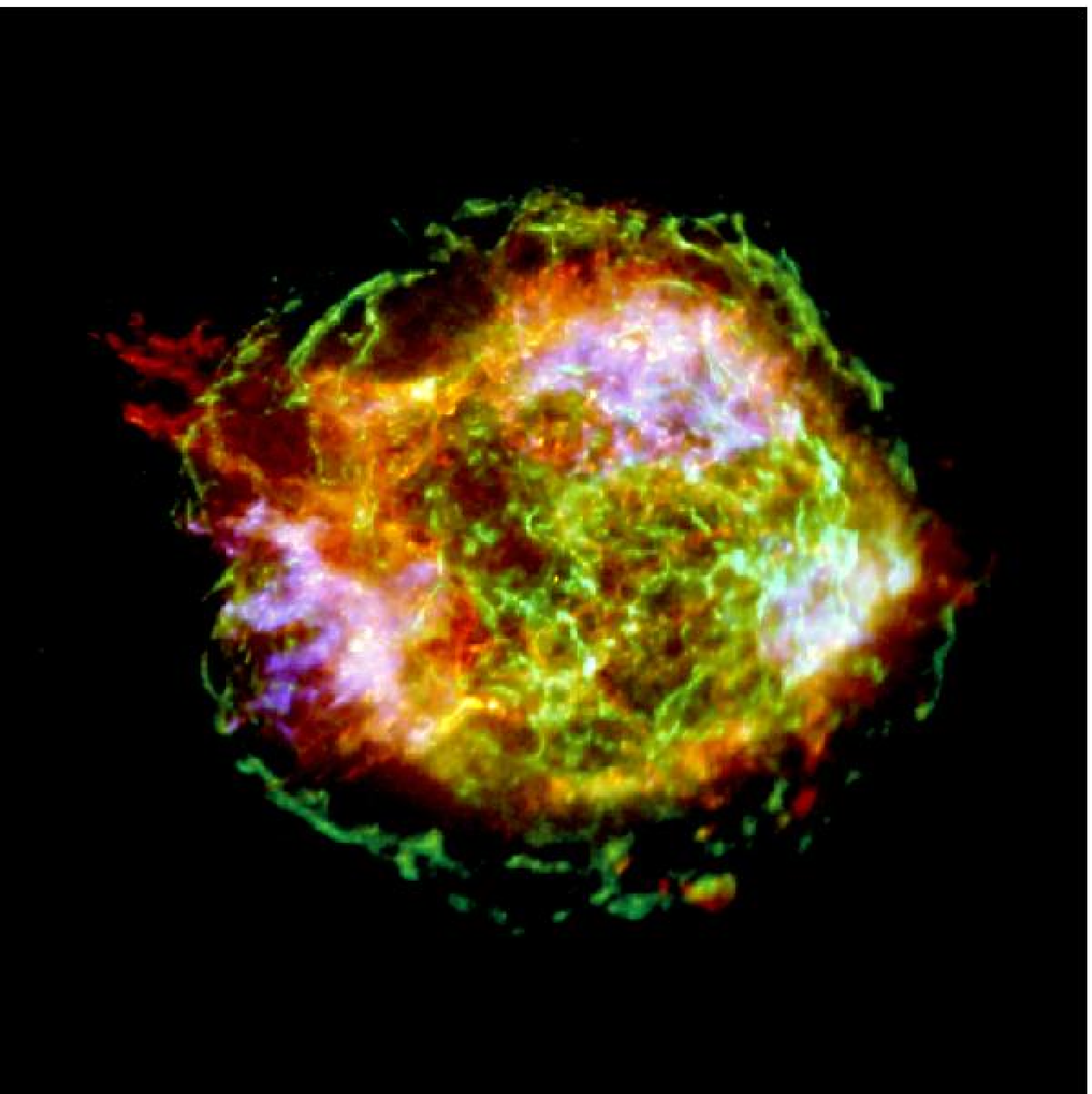,width=0.9\linewidth}
\caption{\it Cas~A in X-ray synchrotron light, observed with
            Chandra. Image courtesy of NASA/CXG/GSFC/U.Hwang et al..}
\label{fig:CasAChandra}
\end{center}
\end{minipage} 
\hspace{0.06\linewidth}
\begin{minipage}[t]{0.47\linewidth}
\begin{center}
\epsfig{file=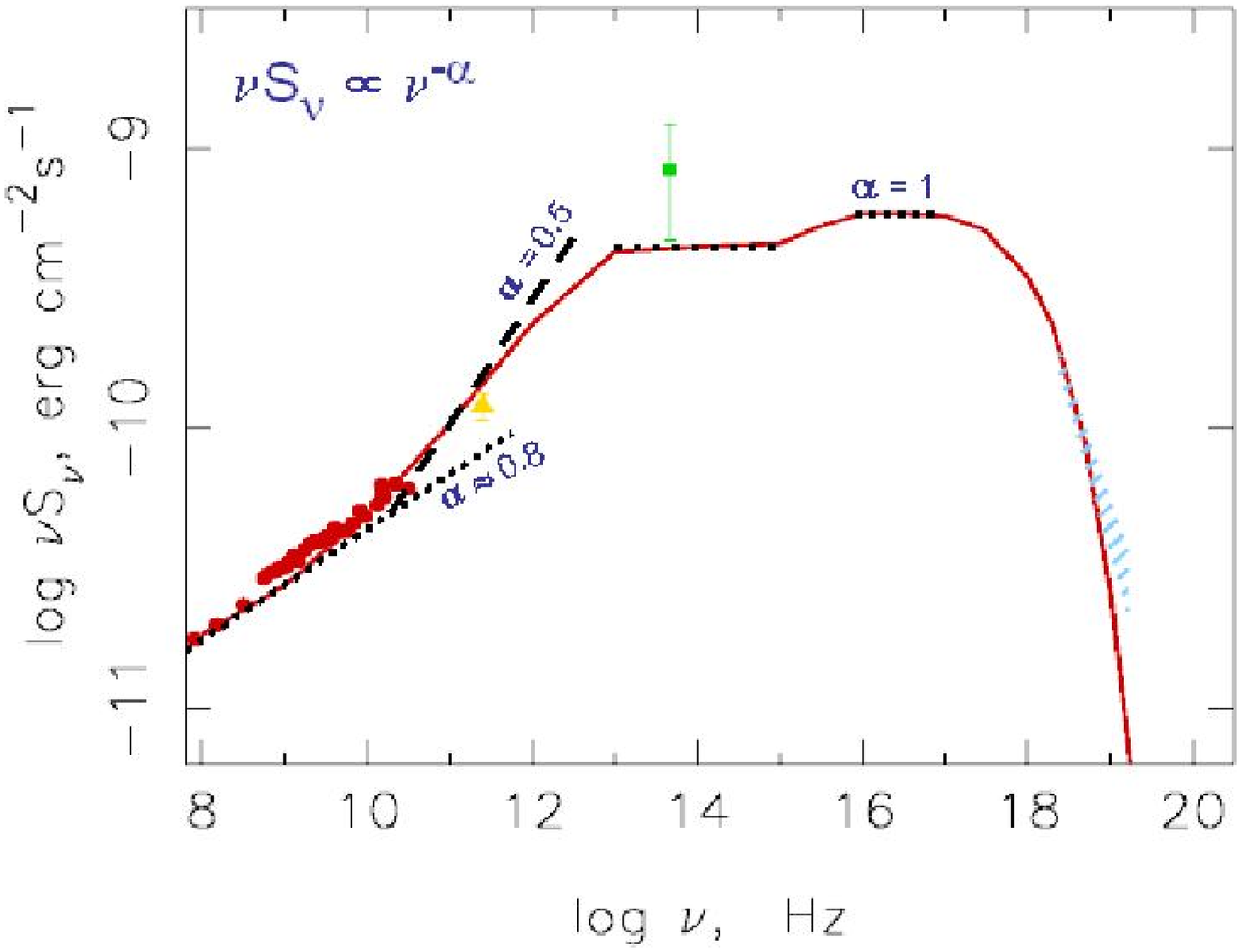,width=1.\linewidth}
\caption{\it Synchrotron spectral energy density (SED) of Cas~A: data and 
  theoretical spectrum \cite{berezhko03a}. The
  low-frequency photon spectrum has a spectral index $\alpha \approx 0.8$
  (dotted curve), not quite increasing to the test particle value $\alpha
  \approx 0.5$ before in the sub-mm region the electrons inside the SNR are
  already radiatively cooled, so that $\alpha \approx 1$ before the cutoff.}
\label{fig:CasASynch}
\end{center}
\end{minipage}
\end{figure}
$\gamma$-ray emission; in addition, we can then determine key theoretical
quantities like the effective magnetic field and the rate of injection of
suprathermal ions into the acceleration process. In this way it becomes
possible to calculate the energy density and the energy spectrum of nuclear CRs
in the SNR from theory. To obtain quantitative results, obviously some further
astronomical multi-wave-length (MWL) information is required. This concerns the
source distance, the angular size/morphology, the SNR expansion rate, and the
circumstellar density structure. SNe also result from two basically different
physical processes whose main manifestations are type Ia explosions
(deflagration/detonation of an accreting White Dwarf) and type II explosions
(core collapse of a massive star), with several variants according to the mass
of the progenitor star. These different SN types are connected with different
magnitudes of the ejected mass. Therefore also the explosion type must be known
from astronomical measurements.

In this talk I will review the general status of the $\gamma$-ray observations,
emphasizing the non-spherical aspects of the nonthermal emission from SNRs and
the relation of theory and experiment in this field. In this light I will
describe four major sources: Cas~A, SN 1006, RX J1713.7-3946, and Vela Jr. I
will end with a short discussion of the contribution to nonthermal SNR research
by the recent Galactic Plane Scan, performed with the {\it H.E.S.S.}
experiment. In some respects this is an extension of a paper given at the 28th
ICRC in Tsukuba \cite{voelk03a}.
% {\em $\backslash$cite\{V\"olk2003Tsukuba\}}.

\section{Gamma-ray detectability of SNRs and their non-spherical aspects}

The {\it EGRET} instrument on the {\it Compton Gamma Ray observatory (CGRO)}
has not been able to find unequivocal evidence for $\gamma$-ray emission from
SNRs in its energy range below a few GeV. The reasons are the low source flux,
even for objects as close as 1 kpc, and the large angular extent of $\sim
1^{\circ}$ -- of the order of the diameter of the full Moon -- which implies a
high $\gamma$-ray background from diffuse Galactic CRs. If we anticipate the
particle energy spectra in the sources to be much harder than the energy
spectra of the diffuse Galactic CRs, then the signal-to-background ratio
decreases with decreasing $\gamma$-ray energy, making the GeV range generally
unfavorable for detection. In contrast, ground-based imaging Cherenkov
telescopes have made several detections in dedicated VHE observations (see
section 4), and the {\it H.E.S.S.} experiment has identified SNR counterparts
for several sources found in its Galactic Plane Survey. The reasons are the
large effective area of these telescopes/telescope systems, the relatively
lower diffuse $\gamma$-ray background at TeV energies, and the much better
angular resolution of $\sim 10^{-1}$~degrees compared to {\it EGRET}. This
makes nearby SNRs at least marginally detectable \cite{drury94}.

\subsection{Non-spherical aspects of SNRs}
Theoretical models for diffusive shock acceleration at SNRs face the difficulty
of having to cope with the fundamentally non-planar and even non-spherical
geometry of a point explosion into an environment that lacks spherical
symmetry. The dynamics is described by kinetic equations for the particle
distributions $f(p,r,t)$ as functions of particle momentum $p$, radial distance
$r$ and time $t$, nonlinearly coupled with the hydrodynamics of the thermal
gas. Only spherically symmetric solutions are available until now which solve
this intrinsically time-dependent problem
\cite{berezhko94,berezhko96}.
It is clear on the other hand that the magnetic field, which regulates the
particle injection rate into the acceleration process, cannot be spherically
symmetric and is even on average at best axially symmetric in SNRs.

The simplest case is a type Ia SN in a uniform interstellar medium and magnetic
field, with SN 1006 as the clearest example.
(Fig.~\ref{fig:ASCA1006}). The time-average magnetic field line geometry is
schematically shown in Fig.~\ref{fig:injection} \cite{voelk03b}.

For kinematic reasons injection of suprathermal ions escaping from a
thermalized downstream region can only occur for quasi-parallel shocks, where
the instantaneous angle $\Theta_{\mbox{nB}} \ll\pi/2$. And clearly acceleration
can occur only at those parts of the shock surface, where particles can be
injected. Particle acceleration is also directly connected with the
self-excitation of Alfv$\acute{v}$en waves which stochastically change
$\Theta_{\mbox{nB}}$. As a consequence we have (i) a stochastic self-limitation
of the ion injection rate $\eta$ through nonlinear wave production, from
$\eta_{\parallel} \approx 10^{-2}$ to an $\eta_{\mbox{eff}} \approx 10^{-4}$,
plus (ii) a systematic reduction of $\eta$ due to the overall average field
morphology, i.e. strong wave production can occur only locally in the polar
regions, (iii) the hadronic $\gamma$-ray emission is therefore also dipolar for
uniform external field $\underline{B}_1$, and (iv) the same is true for the
synchrotron emission as a result of field amplification by factors between 5
and 10 in the ion acceleration regions \cite{bell04,voelk05a,ballet05}, 
with essentially lower emission from the extensive equatorial region. This last
consequence has been impressively proven in a recent analysis of the XMM data
for SN 1006 by \cite{rothenflug04}.

Altogether this injection asymetry requires a reduction of the overall
acceleration efficiency of nuclear particles as calculated in the spherically
symmetric model. The reduction factor is given by the ratio of the polar areas
to the total shock surface area. This ratio is about 0.2 for a case like SN
1006. In order to reach an overall acceleration efficiency of 10\% this
requires the shock regions in which acceleration actually occurs to achieve an
acceleration efficiency of about 50\%. Such a high efficiency implies an
extremely nonlinear acceleration process with a strong backreaction of the
accelerated particles on the shock structure.

\section{Comparison with theory}

The comparison with theory is of course an essential aspect. However, at
present the theory is still incomplete. The full solution of the Fokker-Planck
transport equations for the distribution functions $f(p,r,t)$ of nuclear
particles and electrons, coupled with the hydrodynamics of the thermal plasma
through the CR pressure gradient and wave dissipation, requires even in
spherical symmetry the knowledge of several ``unknowns'': the effective,
amplified magnetic field strength $B_{\mbox{eff}}$, the actual proton injection
rate $\eta_{\mbox{eff}}$, and the amplitude of the electron distribution. These
unknowns can only be determined through an analysis of the synchrotron
observations which involve the relativistic electron component. This analysis
is actually possible, because for particle energies $E \gg m_{\mbox{p}} c^2$,
corresponding to ultra-relativistic protons, electrons behave like protons in
the acceleration process (e.g. \cite{berezhko05}).

 \begin{figure}
\begin{minipage}[t]{0.47\linewidth}
\begin{center}
\epsfig{file=fig5.eps,width=0.9\linewidth}
\caption{\it Data of an individual Chandra 2-10 keV filamentary structure in
Cas~A and model fit to these data, interpreted as the result of strong
postshock synchrotron losses (at $\psi < 0$).}
\label{fig:CasAScale}
\end{center}
\end{minipage} 
\hspace{0.06\linewidth}
\begin{minipage}[t]{0.47\linewidth}
\begin{center}
\epsfig{file=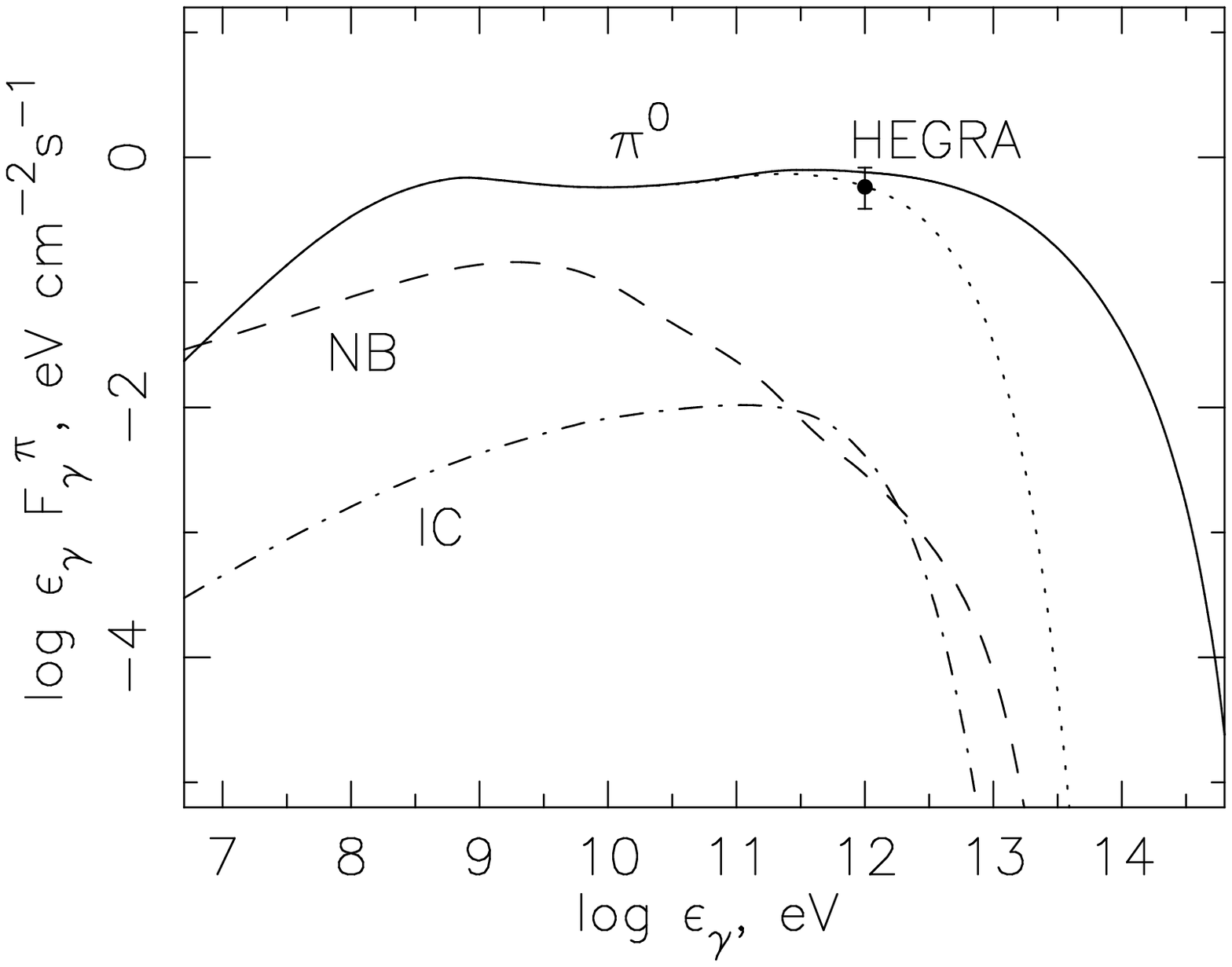,width=0.9\linewidth}
\caption{\it Gamma-ray spectral energy density for Cas~A. The
(unconfirmed) {\it HEGRA} detection and the {\it EGRET} upper limit
 are shown together with the theoretical prediction for
the $\pi^0$-decay ({\it full line}), IC ({\it dashed line}) and NB ({\it
dash-dotted line}) emissions.}
\label{fig:CasAGamma}
\end{center}
\end{minipage}
\end{figure}
I will only summarize the situation here.

The electrons are parasitically accelerated in an environment produced by the
accelerating nuclear particles because they can not modify the shock themselves
due to their small energy density. And therefore at energies $E \gg
m_{\mbox{p}} c^2$ their momentum distribution equals in form that of the
nuclear particles except for radiative (synchrotron) losses:

\begin{itemize}
\item The radio synchrotron spectrum is generally steeper than in the test
particle approximation, because the radiating low-energy electrons ``see'' only
the discontinuous subshock in the thermal gas and not the full shock transition
that includes the extended CR precursor. Interpreting this effect in terms of
nonlinear shock modification by the accelerated nuclear component determines
the ion injection rate $\eta_{\mbox{eff}}$ 

\item The energy of the radio electrons should be substantially lower than 
$10 m_{\mbox{p}}c^2$. Together with the requirement to fit the entire
synchrotron spectrum, including the cutoff at hard X-rays, this determines the
(amplified) magnetic field strength $B_{\mbox{eff}}$

\item The amplitude of the relativistic electron density then follows from 
the amplitude of the synchrotron spectrum, which fixes the electron-to-proton
ratio in the accelerated CRs
\end{itemize}

\section{Individual SNRs}
\subsection{Cas~A}
Cas~A is presumably the result of a so-called Type Ib SN, the core collapse of
a massive Wolf-Rayet star that has already shed its hydrogen envelope through a
fast stellar wind (Fig.~\ref{fig:CasAChandra}). In a detailed model for the
thermal X-ray emission by \cite{borkowski96} the final Wolf-Rayet wind phase
has compressed the inner part of the dense slow wind from the preceeding Red
Supergiant (RSG) phase into a dense luminous shell; in turn the subsequent SNR
shock has already reached the unperturbed RSG wind region beyond the shell.
 
The strong deviation of the synchrotron spectral energy density from a test
particle spectrum below some tens of GHz (Fig.~\ref{fig:CasASynch}) implies a
strong modification of the shock by accelerated nuclear particles, an amplified
post-shock field $B_{\mbox{eff}}\approx 200 \mu$G, and a field
$B_{\mbox{eff}}\approx 500 \mu$G in the RSG wind shell \cite{berezhko03a}.

The X-ray morphology of Cas~A from {\it Chandra} also shows strongly pronounced
filamentary structures of the outer shock (Fig.~\ref{fig:CasAScale}), analyzed
by \cite{vink03} and \cite{berezhko04}. The multi-TeV electrons accelerated at
the shock form a very thin quasi-spherical shell -- thinner than that of a
children's rubber ball -- as the result of violent synchrotron cooling. This
cooling scale determines the interior amplified field. It turns out that this
field ``measurement'' agrees with that using the spectral distortion in the
radio frequency range within the errors of 20 to 30 percent.  Similar results
have been obtained for SN 1006 \cite{berezhko03b,ksenofontov05} and Tycho`s SNR
\cite{voelk05a,ballet05}.

The theoretical prediction of the $\gamma$-ray fluxes
(Fig.~\ref{fig:CasAGamma}, from \cite{berezhko03a}) shows a $\pi^0$-decay
$\gamma$-ray flux that dominates those from IC scattering and NB by two orders
of magnitude at 1 TeV, making it a clear hadronic $\gamma$-ray source by a
large margin. This prediction essentially agrees with the flux detected by the
{\it HEGRA} experiment \cite{aharonian01a}.

The IC and NB flux determinations are quite robust results. A reliable
independent measurement of the $\gamma$-ray flux with the large Northern
Hemisphere telescopes {\it VERITAS} or {\it MAGIC} would therefore be of
paramount importance.

\subsection{SN 1006}
This SNR has been observed in hard X-rays with {\it ASCA} to show purely
nonthermal emission from the two hot spots at the poles, as discussed before,
and this emission was interpreted as synchrotron radiation
\cite{koyama95}. Later high-resolution{\it Chandra} and {\it XMM}
observations strengthened this picture. However, the TeV $\gamma$-ray
detections by the single {\it CANGAROO} telescopes {\it CANGAROO I}
\cite{tanimori98} and {\it CANGAROO II}
\cite{tanimori01} could not be confirmed in a total of 24.5 hours
of observation time by the {\it H.E.S.S.} stereoscopic system
\cite{aharonian05a}. Recent {\it CANGAROO} stereo observations could
not detect the source any more either and have led to the withdrawal of the
earlier $\gamma$-ray detection claims \cite{mori05}.

There are two reasons for the $\gamma$-ray non-detection. First of all, the
magnetic field in the SNR interior is considerably amplified ($B_{\mbox{eff}}
\approx 150 \mu$G), so that the IC radiation is strongly suppressed, given the
observed synchrotron emission. Secondly, the external hydrogen density $N_H$ is
in all probability quite low, $N_H < 0.1$cm$^{-3}$. Since in the Sedov
phase, in which SN 1006 is at present, the $\pi^0$-decay $\gamma$-ray flux
$F_{\gamma}$ is proportional to $N_H^2$, the low gas density implies a low
hadronic $\gamma$-ray emission as well. This situation has been analyzed in
detail by \cite{ksenofontov05}. Given the lowest value of $N_H = 0.05~
\mbox{cm}^{-3}$, discussed in the literature, we expect the $\gamma$-ray
emission to be only a factor of 3 smaller than the present {\it H.E.S.S.} upper
limit for the northeastern polar cap 
\begin{figure}
\begin{minipage}[t]{0.47\linewidth}
\begin{center}
\epsfig{file=fig7.eps,width=0.9\linewidth}
\caption{\it Integral VHE photon flux from the northeastern polar cap of SN
1006. The value $B_0 = 30 \mu$G corresponds to an interior effective field
strength of $150 \mu$G. The CANGAROO and H.E.S.S. data are shown together with
the theoretical flux estimates in \cite{ksenofontov05}.}
\label{fig:SN1006gammaflux}
\end{center}
\end{minipage} 
\hspace{0.06\linewidth}
\begin{minipage}[t]{0.47\linewidth}
\begin{center}
\epsfig{file=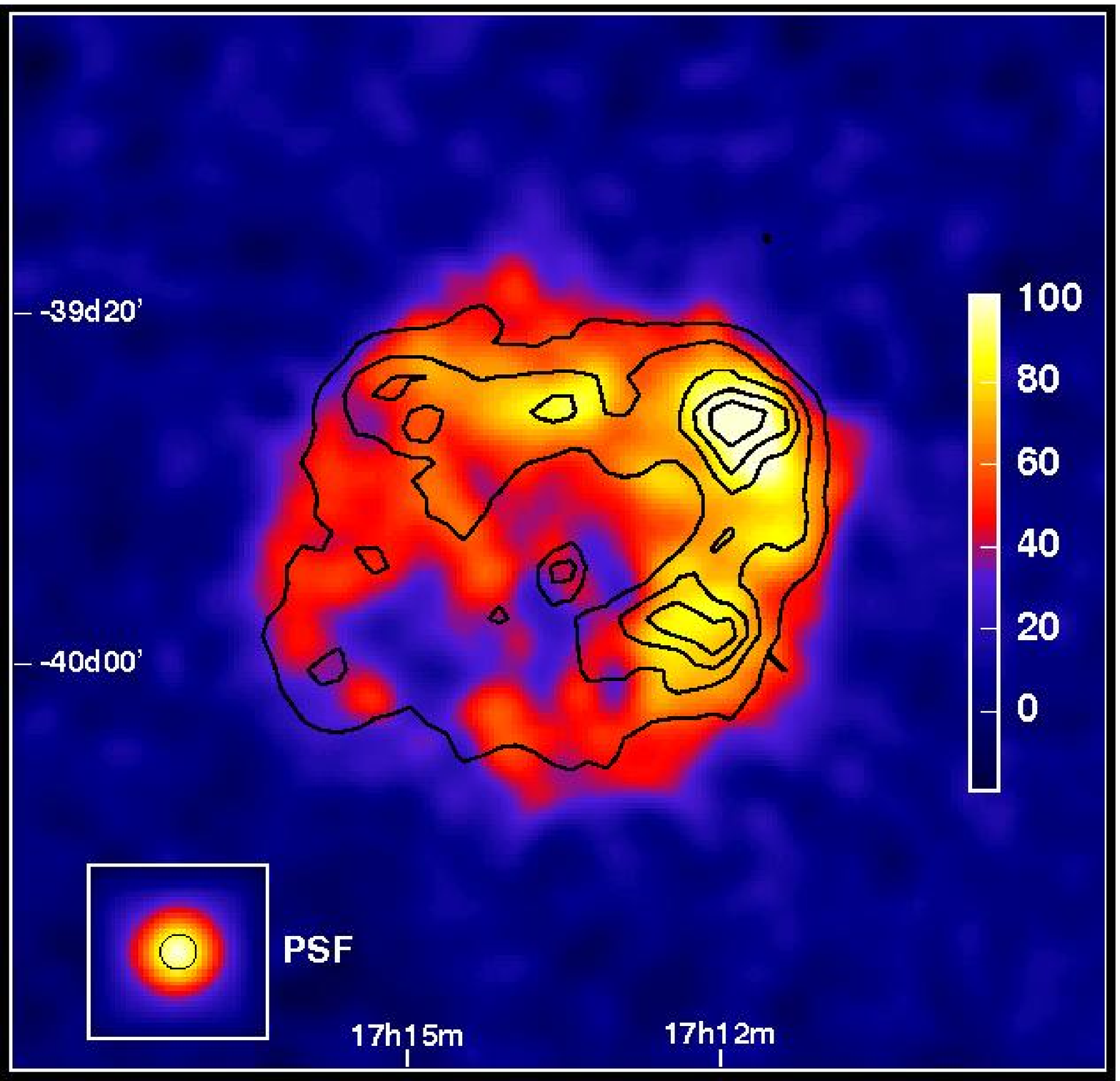,width=0.9\linewidth}
\caption{\it H.E.S.S. image of RX J1713.7-3946 at $\gamma$-ray energies > 800
  GeV, with the ASCA 1 -- 3 keV X-ray contours
  superposed \cite{berge05}.}
\label{fig:RXJimage}
\end{center}
\end{minipage}
\end{figure}
(Fig.~\ref{fig:SN1006gammaflux}). Since SN 1006 is the simplest case of a SNR
with strong nonthermal emission in radio and X-rays and therefore the
theoretically best understood object of its kind, it would obviously be
important to detect it also in TeV $\gamma$-rays in a deep observation of about
200 hours with {\it H.E.S.S.} or {\it CANGAROO}.

\subsection{SNR RX J1713.7-3946}
Originally found in the {\it ROSAT} X-ray survey
\cite{pfeffermann96}, SNR RX J1713.7-3946 was detected at VHE
energies by {\it CANGAROO} \cite{muraishi00} and interpreted as an
IC-dominated TeV source. {\it ASCA} observations \cite{koyama97,slane99}
had shown that the X-ray spectrum is a purely
nonthermal continuum. In 2002 the {\it CANGAROO} group revised its
interpretation and rather favored a hadronic scenario from the shape of the TeV
energy spectrum \cite{enomoto02}. This phenmenological
interpretation was questioned by \cite{reimer02} and
\cite{butt02} on the grounds that the upper limit from a nearby {\it
EGRET} source 3EG 1714-3857, believed to be associated with the TeV source, was
inconsistent with the hadronic extrapolation of the {\it CANGAROO} spectrum to
the GeV region. Such different views were not entirely surprising at the time,
considering the complex morphology of the general region in which the source is
embeded. This is particularly well visible in the CO map. Nevertheless, at the
present time CO data \cite{fukui03} give a most probable kinematic
source distance of 1 kpc.
% In addition the radio emission is
% rather faint, measured only at two nearby frequencies up to now, so that
% there is considerable freedom to construct the synchrotron spectrum.

The {\it H.E.S.S.} experiment subsequently confirmed the {\it CANGAROO}
detection and could give for the first time a spatially resolved VHE image of a
SNR \cite{aharonian04}, whose overall shell structure correlated
closely with the shell structure in {\it ASCA} 1--3 keV X-rays. The remnant
diameter is about $1^{\circ}$. This was unambiguous proof for the acceleration
of charged particles to energies beyond 100 TeV\footnote{The following
  discussion of this source and the figures \ref{fig:RXJimage} and
  \ref{fig:RXJspectrum} are based on the more recent H.E.S.S. results, not yet
  released at the time of the conference.}

Apart from the fit shown in Fig.~\ref{fig:RXJspectrum}, the H.E.S.S. VHE
differential spectrum can be equally well fitted by a power law with
exponential cutoff $\propto E^{-\Gamma} \exp^{-E/E_c}$, with $\Gamma = 1.98 \pm
0.05$ and $E_c = 12 \pm 2$. At energies $E \ll E_c$ this measured spectrum
corresponds to the test particle limit of diffusive shock acceleration theory
for nuclear particles in a strong shock. Even ignoring nonlinear backreaction
effects the extrapolation of the charged particle spectrum to lower energies
with a proton spectrum $\propto E^{-2}$ gives a hadronic $\gamma$-ray spectrum
below an improved upper limit of {\it EGRET} which now assumes that RX
J1713.7-3946 is not linked to the known {\it EGRET} source 3EG 1714-3875
\cite{aharonian05b}.

Making in fact the best case for an IC interpretation by assuming a very low
magnetic field strength of about $10 \mu$G inside the SNR, the resulting IC
spectrum fits the {\it H.E.S.S.} data quite poorly. This outweighs the good
correlation between the X-ray synchrotron and the $\gamma$-ray emissions which
at first sight would suggest a leptonic origin of the $\gamma$-ray emission as
well. In addition, already a small amplification of the magnetic field in the
remnant rules out a dominant IC emission and a fortiori a dominant NB, whereas
a hadronic $\gamma$-ray spectrum continues to fit the data quite well
\cite{aharonian05b}. Despite the complex CO morphology and unknown
age of this remnant I believe that a hadronic interpretation of the
$\gamma$-ray emission from RX J1713.7-3946 is clearly favored. The most
plausible acceleration scenario is that of a massive progenitor star which,
over its long evolution time, had produced a large stellar wind bubble into
which finally the SN exploded about a thousand years ago (see e.g. Fig. 5 of
\cite{berezhko00}). Only this makes the almost circular X-ray
morphology understandable that was found by ASCA and XMM
\cite{cassam04,hiraga05}. And it can explain the
low effective density inside the SNR \cite{cassam04}. A detailed
theoretical model is needed to investigate the consistency of such a picture
with the existing MWL evidence.

\subsection{Vela Jr.}

\begin{figure}
\begin{minipage}[t]{0.47\linewidth}
\begin{center}
\epsfig{file=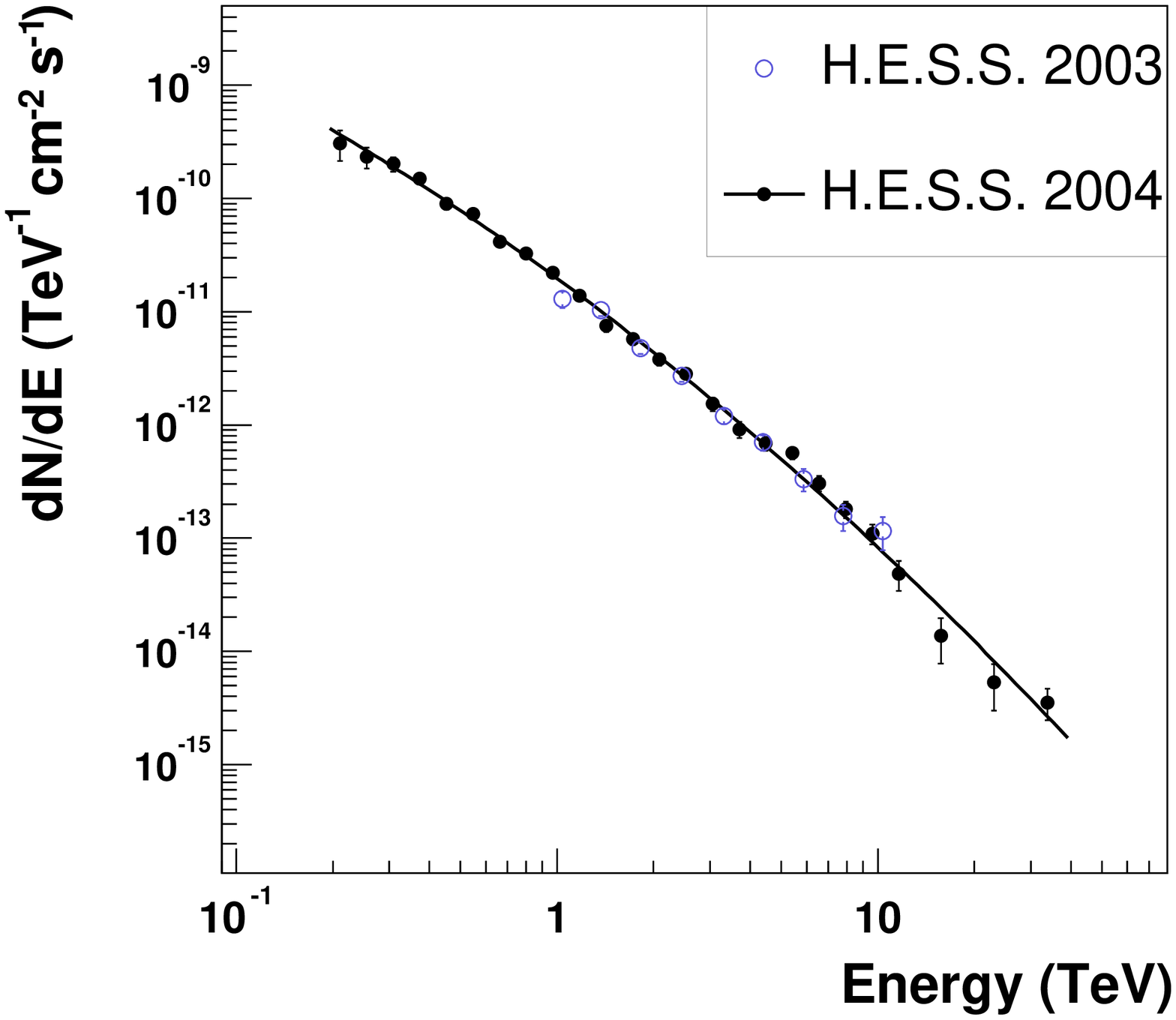,width=0.9\linewidth}
\caption{\it H.E.S.S. differential $\gamma$-ray spectrum of RX J1713.7-3946 for
  the whole region of the SNR {\it solid black circles}. The best fit of a
  power law with energy-dependent photon index is plotted as a {\it black
    line}. The H.E.S.S. 2003 data are given by the {\it blue open circles}.
  Error bars are $\pm 1\sigma$ statistical errors
  \cite{aharonian05b}.}
\label{fig:RXJspectrum}
\end{center}
\end{minipage} 
\hspace{0.06\linewidth}
\begin{minipage}[t]{0.47\linewidth}
\begin{center}
\epsfig{file=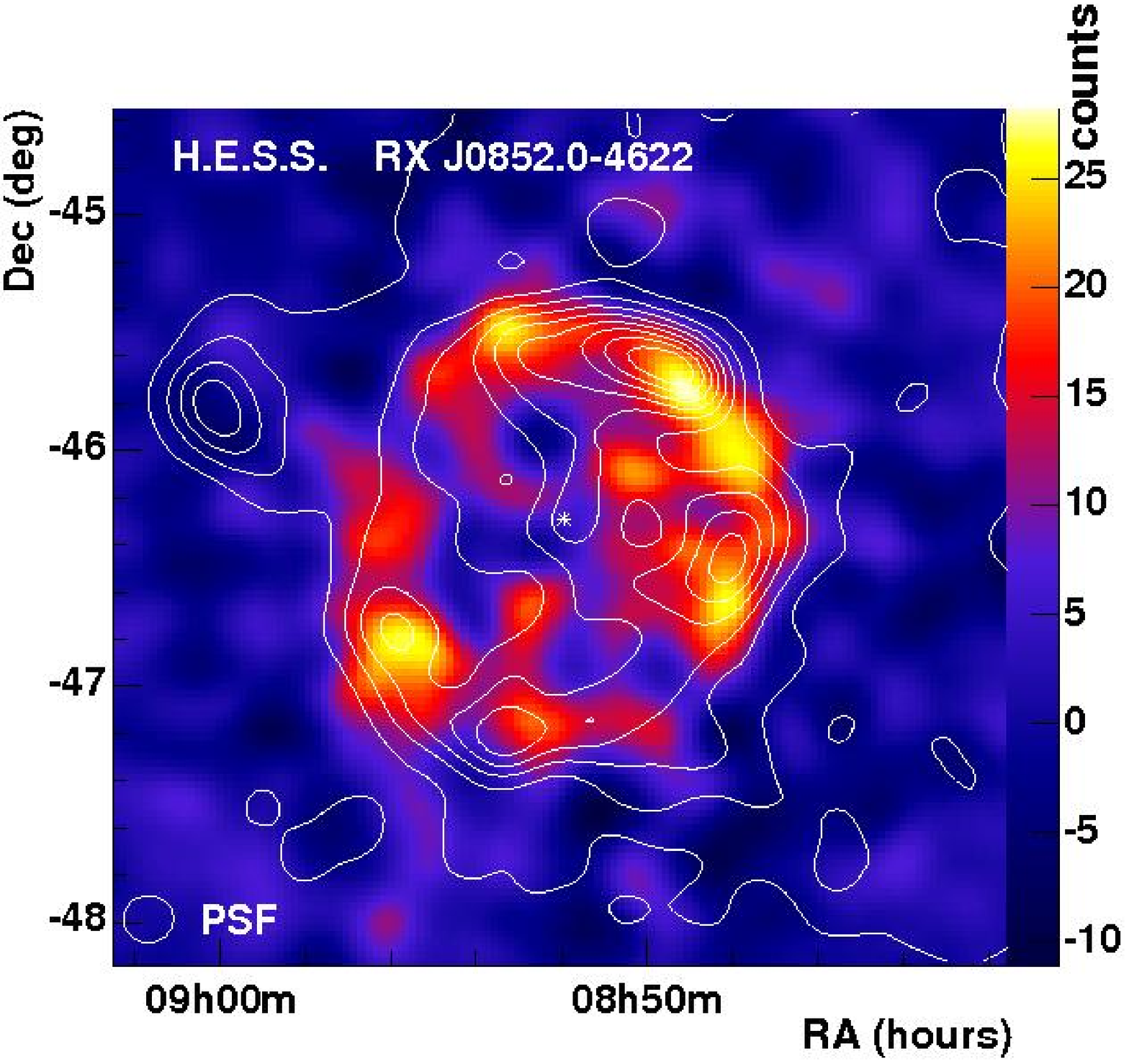,width=0.9\linewidth}
\caption{\it Vela Jr. in TeV $\gamma$-rays. The color scale is in number of
$\gamma$-ray events. The total significance of this 3.2 hours lifetime
observation with {\it H.E.S.S.} is $12 \sigma$. The source radius is four times
the radius of the full Moon.}
\label{fig:VelaJrimage}
\end{center}
\end{minipage}
\end{figure}

The SNR RX J0852.0-4622, also called Vela Jr., was also found with {\it ROSAT}
\cite{aschenbach98} as a very large $2^{\circ}\times 2^{\circ}$
quasi-circular X-ray structure and later confirmed by {\it ASCA}
\cite{tsunemi00,slane01}.
In the VHE range it was detected by {\it CANGAROO} \cite{katagiri05}
and by {\it H.E.S.S.} \cite{aharonian05c}, see
Fig.~\ref{fig:VelaJrimage}.

The flux above $1~TeV$ equals 1.4 times the flux from RX J1713.7-3946 and is
about equal to the flux from the Crab Nebula. The differential spectrum is a
hard power law spectrum with $\Gamma = 2.1 \pm 0.1 \pm 0.1$
\cite{aharonian05c}.
 
The physical characteristics of Vela Jr. appear to be rather similar to those
of RX J1713.7-3946, even though the source may be even closer and younger.
Interestingly, also a narrow filament was recently found in {\it Chandra} data
\cite{bamba05}.  If we interpret this filament as part of the outer
SNR shock its sharpness suggests magnetic fields at least in the $100 \mu$G
range, almost independently of the detailed astronomical properties of the
source \cite{voelk05b}. Therefore again a hadronic interpretation
of the $\gamma$-ray emission is plausible, even though little theoretical
analysis has been performed up to now. Further and more detailed results on the
morphology and spectrum are expected to be available soon. They should make
this source the second major VHE SNR source in the Southern Hemisphere.

\section{Galactic Plane Scan SNRs, Conclusions}

The {\it H.E.S.S.} Galactic Plane Scan \cite{aharonian05d,aharonian05e}
has revealed some 20 odd new sources until now.
Several of these soures can be associated with SNRs with reasonable certainty,
but many of them are still un-identified at the present time. It is clear that
further study of these SNRs is required, both at VHE energies and at other
wavelengths. An example is HESS J1813-178 where, quickly following the initial
H.E.S.S. publication of the first scan results, a SNR was found in existing
{\it VLA} radio data \cite{brogan05}, and a known coincident {\it
  ASCA} source was found also by {\it INTEGRAL} in the 20-100 keV band
\cite{ubertini05}.

In the present context such studies are important to identify the statistical
VHE properties of the Galactic SNR and thus to control whether the few very
nearby objects, that can be studied in great detail, are typical
representatives of this population or not.

Even though it is difficult to make such population studies in the Northern
Hemisphere, there are at least two bright historical SNRs in this Hemisphere,
Cas~A and Tycho's SNR. Cas~A has been detected by {\it HEGRA} in TeV
$\gamma$-rays with rather low significance, whereas the {\it HEGRA} flux upper
limit for Tycho's SNR \cite{aharonian01b}
is only a factor of unity above the theoretical VHE prediction
\cite{voelk02,voelk05a}. Together with other known
Northern Hemisphere SNRs, the confirmation/detection -- or not -- of these
objects with {\it VERITAS} or {\it MAGIC} would be of high significance for the
question of CR origin in SNRs.

\section*{Acknowledgments}
I would like to thank the members of the H.E.S.S. collaboration for discussions
about the {\it H.E.S.S.} and {\it HEGRA} results mentioned in this talk. I also
gratefully acknowledge the collaboration with E.G. Berezhko and
L.T. Ksenofontov on the theory of SNRs over the years.

% Files should be sent by the {\bf 3}$^{\rm rd}$ {\bf of October 2005}
% to: \\
% \centerline {\tt
% cherenkov2005.proceedings@poly.in2p3.fr}
% with the keyword ``Cherenkov 2005 Proceedings'' in the subject. 
% Please, make sure that eps files you are sending are {\it true eps
% files} and can be used without any problem on all platforms!

%%%%%%%%%%%%%%%%%%%%%%%%%%%%%%%%%%%%%
% Label to flag the last page of your contribution
% Replace Yourname by your name starting with a capital letter
%
\label{VoelkEnd}
 

\begin{thebibliography}{99}

%%
%%  bibliographic items can be constructed using the LaTeX format in SPIRES:
%%    see    http://www.slac.stanford.edu/spires/hep/latex.html
%%  SPIRES will also supply the CITATION line information; please include it.
%%

\bibitem{hillas05}   A.M. Hillas, {\it J. Phys. G} {\bf 31}, R95 (2005)


\bibitem{koyama95}   K. Koyama et al., {\it Nature} {\bf 378}, 225 (1995)
  

\bibitem{voelk97}   H.J. V\"olk, {\it Proc. ``Towards a Major Atmospheric
    Cherenkov Detector-V, Kruger Nat'l Park}, ed. O.C. de Jager,
  Westprint-Potchefstrom, RSA, p. 87 ff (1997)


\bibitem{voelk03a}   H.J. V\"olk, {\it Frontiers of Cosmic Ray Science,
    Proc. 28th ICRC} (Tsukuba) {\bf 8}, 29 (2003)
  

\bibitem{drury94}  L.O'C. Drury et al., {\it Astron. Astrophys.} {\bf 287}, 959
  (1994)


\bibitem{berezhko94}  E. Berezhko et al., {\it Astroparticle Phys.} {\bf 2}, 215 (1994)


\bibitem{berezhko96}  E. Berezhko et al., {\it J. Exp. Theor. Phys.} {\bf 82}, 1 (1996)


\bibitem{voelk03b}  H.J. V\"olk et al., {\it Astron. Astrophys.} {\bf 409}, 563
  (2003)
  
\bibitem{voelk05a} H.J. V\"olk et al., {\it Astron. Astrophys.} {\bf 433}, 2929
  (2005)
  
\bibitem{ballet05} J. Ballet, {\it Adv. Space Res.} {\bf 35}, (2005); 
                   (astro-ph/0503309)  


\bibitem{rothenflug04}  R. Rothenflug et al., {\it Astron. Astrophys.} {\bf
    425}, 121 (2004)


\bibitem{bell04}  A.R. Bell, {\it MNRAS} {\bf 182}, 550
  (2004)

\bibitem{berezhko05} E.G. Berezhko, {\it Adv. Space Res.} {\bf 35}, 1031 (2005)


\bibitem{borkowski96}  K.J. Borkowsky et al., {\it Astrophys. J} {\bf 466}, 866
  (1996)

  
\bibitem{berezhko03a} E.G. Berezhko et al., {\it Astron. Astrophys.} {\bf 400},
  971 (2003)


\bibitem{vink03}  J. Vink et al., {\it Astrophys. J.} {\bf 548}, 758
  (2003)

  
\bibitem{berezhko04} E.G. Berezhko et al., {\it Astron. Astrophys.} {\bf 419},
  L27 (2004)
  
\bibitem{berezhko03b} E.G. Berezhko et al., {\it Astron. Astrophys.} {\bf 412},
  L11 (2003)

\bibitem{ksenofontov05}  L.T. Ksenofontov et al., {\it Astron. Astrophys.} {\bf
  443}, 973 (2005); (astro-ph/0508318)


\bibitem{berge05} D. Berge et al. (for the H.E.S.S. collaboration), {\it
    Proc. 29th ICRC} Pune (2005), {\bf 4}, 117

  
\bibitem{aharonian01a} F.A. Aharonian et al., {\it Astron. Astrophys.} {\bf
    370}, 112 (2001)


\bibitem{tanimori98}  T. Tanimori et al., {\it Astrophys. J.} {\bf 497}, L25
(1998)


\bibitem{tanimori01} T. Tanimori et al., {\it Proc. 27th ICRC} (Hamburg) {\bf
    6}, 2465 (2001)


\bibitem{aharonian05a} F.A. Aharonian et al., {\it Astron. Astrophys.} {\bf
    437}, 95 (2005)


\bibitem{mori05}  M. Mori, {\it these proceedings}, p. 19 (2005)
  
  
\bibitem{pfeffermann96} E. Pfeffermann et al., in {\it ``Roentgenstrahlung from
    the Universe''}, 267 (1996)

\bibitem{muraishi00} H. Muraishi et al., {\it Astron. Astrophys.} {\bf 354},
  L57 (2000)

\bibitem{koyama97}  K. Koyama et al., {\it PASJ} {\bf 49}, L7 (1997)


\bibitem{slane99}  P. Slane et al., {\it Astrophys. J.} {\bf 525}, 357 (1999)


\bibitem{enomoto02}  R. Enomoto et al., {\it Nature} {\bf 416}, 823 (2002)

  
\bibitem{reimer02} O. Reimer et al., {\it Astron. Astrophys.} {\bf 390}, L43
  (2002)


\bibitem{butt02}   Y.M. Butt et al., {\it Nature} {\bf 418}, 499 (2002)


\bibitem{fukui03}  Y. Fukui et al., {\it PASJ} {\bf 55}, L61 (2003)


\bibitem{aharonian04}  F.A. Aharonian et al., {\it Nature} {\bf 432}, 75 (2004)


\bibitem{aharonian05b} F.A. Aharonian et al., {\it Astron. Astrophys.} in press
  (2005); arXiv:astro-ph/0511678


\bibitem{berezhko00}  E.G. Berezhko et al., {\it Astron. Astrophys.} {\bf
    357}, 283 (2000)


\bibitem{cassam04} G. Cassam-Chena\"i et al., {\it Astron. Astrophys.} {\bf 427},
  199 (2004)

\bibitem{hiraga05} J.S. Hiraga et al., {\it Astron. Astrophys.} {\bf 431}, 953 (2005)


\bibitem{aschenbach98} B. Aschenbach, {\it Nature} {\bf 396}, 141 (1998)


\bibitem{tsunemi00}  H. Tsunemi et al., {\it PASJ} {\bf 52}, 887 (2000)


\bibitem{slane01}  P. Slane et al., {\it Astrophys. J.} {\bf 548}, 814 (2001)

  
\bibitem{katagiri05} H. Katagiri et al., {\it Astrophys. J.} {\bf 619}, L163
  (2005)


\bibitem{aharonian05c}  F.A. Aharonian et al., {\it Astron. Astrophys.} {\bf
    437}, L7 (2005)


\bibitem{bamba05}  A. Bamba et al., {\it Astrophys. J.} {\bf 632}, 294 (2005)
  
\bibitem{voelk05b} H.J. V\"olk et al., {\it Proc. 29th ICRC} Pune (2005)
{\bf 3}, 233


\bibitem{aharonian05d} F.A. Aharonian et al., {\it Science} {\bf 307}, 1938
(2005)


\bibitem{aharonian05e} F.A. Aharonian et al., {\it Astrophys. J.}, to appear
(2005)


\bibitem{brogan05} C.L. Brogan et al., {\it Astrophys. J.} {\bf 629}, L105
(2005)
 

\bibitem{ubertini05} P. Ubertini et al., {\it Astrophys. J.} {\bf 629}, L1009
(2005)
 

\bibitem{aharonian01b}  F.A. Aharonian et al., {\it Astron. Astrophys.}
  {\bf373}, 292 (2001)
 
\bibitem{voelk02} H.J. V\"olk et al., {\it Astron. Astrophys.} {\bf 396}, 649
  (2002)
 




% . Yourname, ``Instruction for producing
%   Cherenkov 2005 proceedings'', {\it these proceedings
%   pp.~\pageref{YournameStart}-\pa geref{YournameEnd}}



\end{thebibliography}
\end{document}